\documentclass[12pt,epsf,amssymb,ulem,qsymbols]{article}
\usepackage{tabularx}
\usepackage{array}
\usepackage{graphics}
\usepackage{graphicx}
\usepackage{psfrag}
\usepackage{epsfig}
\usepackage{amsmath}
\usepackage{amssymb}
\usepackage[normalem]{ulem}
\usepackage{setspace}
\usepackage{rotating}
\usepackage{colortbl}
\usepackage{tabularx}
\usepackage{longtable}
\usepackage{slashed}
\makeatletter
\usepackage{textcomp}
\usepackage[usenames,dvipsnames]{xcolor}
\usepackage{relsize}

\usepackage{verbatim}
\usepackage{cite}

\setlongtables

\newcommand{\bea}{\begin{eqnarray}}
\newcommand{\eea}{\end{eqnarray}}
\newcommand{\beq}{\begin{equation}}
\newcommand{\eeq}{\end{equation}}
\newcommand{\ec}{\end{center}}
\newcommand{\bc}{\begin{center}}

\newcommand{\gev}{{\rm GeV}}

\newcommand{\pdir}{p\kern -5.2pt\raise 0.2ex\hbox {/}}

\newcommand{\vdir}{v\kern -5.75pt\raise 0.15ex\hbox {/}}
\newcommand{\kdir}{k\kern -5.75pt\raise 0.15ex\hbox {/}}
\newcommand{\epsdir}{\epsilon\kern -5.0pt\raise 0.15ex\hbox {/}}
\newcommand{\bvdir}{\bar{v}\kern -5.75pt\raise 0.15ex\hbox {/}}
\newcommand{\Ddir}{D\kern -7.75pt\raise 0.20ex\hbox {/}}
\newcommand{\Adir}{A\kern -7.75pt\raise 0.20ex\hbox {/}}
\newcommand{\ldir}{l\kern -5.0pt\raise 0.2ex\hbox{/}}
\newcommand{\varepsdir}{\varepsilon\kern -5.5pt\raise 0.15ex\hbox{/}}

\newcommand{\nn}{\nonumber}

\makeatother

\begin{document}
\unitlength = 1mm

\begin{flushright}
\begin{tabular}{l}
{\tt \footnotesize LPT 15-100}\\
\end{tabular}
\end{flushright}

\begin{center}
\vskip 1.4cm\par
{\par\centering \textbf{ 
\Large \bf Can the new resonance at LHC }}\\
\vskip .35cm\par
{\par\centering \textbf{ 
\Large \bf 
 be a CP-Odd Higgs boson?}}\\
\vskip 1.05cm\par
{\scalebox{.81}{\par\centering \large  
\sc D. Be\v{c}irevi\'c$^a$, E. Bertuzzo$^{b}$, O. Sumensari$^{a,b}$ and R. Zukanovich Funchal$^{b}$}
{\par\centering \vskip 0.65 cm\par}
{\sl 
$^a$~Laboratoire de Physique Th\'eorique (B\^at.~210)\\
Universit\'e Paris Sud and CNRS (UMR 8627), F-91405 Orsay-Cedex, France.}\\
{\par\centering \vskip 0.25 cm\par}
{\sl 
$^b$~Instituto de F\'isica, Universidade de S\~ao Paulo, \\
 C.P. 66.318, 05315-970 S\~ao Paulo, Brazil.}\\
}
\end{center}

\vskip 0.55cm
\begin{abstract}
  A plausible explanation of the recent experimental indication of a resonance
  in the two-photon spectrum at LHC is that it corresponds to the CP-odd Higgs
  boson.  We explore such a possibility in a generic framework of the two
  Higgs doublet models (2HDM), and combine $m_A \approx 750$~GeV with the
  known $m_h =125.7(4)$~GeV to show that the charged Higgs boson and the other
  CP-even scalar masses become bounded from bellow and from above. We show
  that this possibility is also consistent with the electroweak precision data
  and the low energy observables, which we test in a few leptonic and
  semileptonic decay modes.
\end{abstract}
\setcounter{footnote}{0}
\setcounter{equation}{0}
\noindent

\renewcommand{\thefootnote}{\arabic{footnote}}

\setcounter{footnote}{0}
\section{\label{sec-0}Introduction}
In addition to the Higgs boson, $m_h=125.7(4)$~GeV~\cite{PDG}, the experiments
at LHC recently indicated a possibility for a resonance in the diphoton
spectrum at about $750$~GeV~\cite{LHCnew}.  While its spin must be either
$J=0$ or $2$, its parity cannot be yet assessed. If, after improving
statistics and further examining systematics of the data sample, this signal
remains as such, a plausible explanations for the newly observed state could
be the ones offered in
refs.~\cite{Franceschini:2015kwy,Harigaya:2015ezk,Higaki:2015jag,Buttazzo:2015txu,Ellis:2015oso,Angelescu:2015uiz,Bellazzini:2015nxw,DiChiara:2015vdm,Knapen:2015dap,Mambrini:2015wyu,Dev:2014yca}.

The simplest possibility is to consider scenarios with two Higgs doublets
(2HDM) in which the spectrum of scalars consists of two CP-even ($h$ and $H$),
one CP-odd ($A$) and one charged Higgs state ($H^\pm$).  In this paper, we
focus on the possibility of the new state being the CP-odd Higgs and find that
the general theoretical constraints combined with the two known masses result
in the bounds on the remaining two Higgs boson states.  We also show that the
resulting bounds and the proposed scenario satisfy the electroweak precision
tests, and do not significantly modify the low energy (semi-)leptonic decay
modes.

Let us stress that, in a pure 2HDM, the production cross section for the heavy spin-$0$ states 
seems to be too low to explain the claimed excess of $\sigma(gg\to X)B(X\to \gamma\gamma)$, where $X$ stands for the new $750$~GeV resonance, in such a way 
that an extended particle content might be needed~\cite{Angelescu:2015uiz}. However, our conclusions on the spectrum of the model are unlikely to be significantly affected by the additional particles as long as they are fermions.~\footnote{\label{footnote:EWPM} In principle, 
our analysis of the electroweak precision tests would be affected by the presence of new fermions. However, 
the final outcome of such tests would be model dependent.}
One should be cautious and study carefully the signal strength, including the background contamination in the signal region as well as the possible signal-background interference, which in general are model dependent~\cite{Greiner:2015ixr}. It should be reemphasized that other possibilities can be envisaged, such as the one in which $H$ and $A$ are mass degenerate. In this paper, we focus on the possibility of $A$ being the desired state hinted at about $750$~GeV.

\section{\label{sec:2}General Constraints on 2HDM and the Spectrum of Higgses}

The most general CP-conserving 2HDM potential compatible with gauge symmetries of the
Standard Model is given by (see eg. ref.~\cite{Branco:2011iw}),
\begin{align}
\label{v2hdm}
V(\Phi_1,\Phi_2)=m_{11}^2\Phi_1^\dagger\Phi_1 &+m_{22}^2\Phi_2^\dagger\Phi_2-m_{12}^2(\Phi_1^\dagger\Phi_2+\Phi_2^\dagger\Phi_1)+\dfrac{\lambda_1}{2}(\Phi_1^\dagger \Phi_1)^2+\dfrac{\lambda_2}{2}(\Phi_2^\dagger \Phi_2)^2\\
&+\lambda_3 \Phi_1^\dagger\Phi_1 \Phi_2^\dagger\Phi_2+\lambda_4 \Phi_1^\dagger\Phi_2 \Phi_2^\dagger\Phi_1+\dfrac{\lambda_5}{2}\left[ (\Phi_1^\dagger\Phi_2)^2+(\Phi_2^\dagger\Phi_1)^2\right]\nonumber,
\end{align}
where $\Phi_1$ and $\Phi_2$ are the two complex scalar $SU(2)$ doublets with hypercharge $Y=+1$. In the above expression the $\mathcal{Z}_2$ symmetry ($\Phi_{1,2}\to -\Phi_{1,2}$) has been tacitly assumed, except for 
the soft symmetry breaking term proportional to $m_{12}^2$. Assuming that each doublet carries a non-zero vacuum expectation value (vev) one can write,
\begin{equation}
\label{higgs}
\Phi_a 	=
	\begin{pmatrix}
		\phi_a^+\\  \frac{1}{\sqrt{2}}(v_a+\rho_a+i\eta_a) 
	\end{pmatrix}, \qquad a=1,2,
\end{equation}
with both vev's $v_{1,2}$ being associated with the neutral components to avoid a problem of breaking the $U(1)$ symmetry of electromagnetism. A further assumption is the conservation of CP-symmetry in the Higgs sector which translates to $v_{1,2} \in \mathbb{R}$. Two of the six fields ($\phi_{1,2}^+$, $\rho_{1,2}$, $\eta_{1,2}$) are Goldstone bosons and can be gauged away, which then leaves us with the physical spectrum consisting of one charged $H^\pm$, two CP-even neutral $h,H$, and one CP-odd neutral $A$ boson, that are linear combinations of the above fields, namely,
\begin{align}
 &H^+=\phi_1^+\sin\beta -\phi_2^+\cos\beta , &A  =\eta_1\sin\beta-\eta_2\cos\beta , \nn \\
 &H =-\rho_1\cos\alpha -\rho_2\sin\alpha , &h=\rho_1\sin\alpha -\rho_2\cos\alpha,
\end{align}
with $\alpha$ and $\beta$ associated with rotations that diagonalize the mass matrices. Written in terms of parameters given in $V(\Phi_1,\Phi_2)$ one gets, 
\begin{equation}
	 \tan\beta = \dfrac{v_2}{v_1},\qquad\tan(2\alpha)=\dfrac{2(-m_{12}^2+\lambda_{345}v_1 v_2)}{m_{12}^2(v_2/v_1-v_1/v_2)+\lambda_1 v_1^2-\lambda_2 v_2^2},
\end{equation}
where $\lambda_{345}\equiv\lambda_3+\lambda_4+\lambda_5$. After setting $\sqrt{v_1^2+v_2^2}\equiv v^{\rm SM}=246.2$~GeV (which in the following will be referred to as $v$), $\tan\beta$ becomes the free model 
parameter and the quartic couplings $\lambda_{1-5}$ can be expressed in terms of scalar masses and mixing angles as~\cite{Kanemura:2004mg}:
\begin{align}
\lambda_1  &= \frac{1}{v^2}\left(-\tan^2\beta M^2+\frac{\sin^2\alpha}{\cos^2\beta} m_h^2+\frac{\cos^2\alpha}{\cos^2\beta}m_H^2\right), \cr
\lambda_2  &=\frac{1}{v^2}\left(-\cot^2\beta M^2+\frac{\cos^2\alpha}{\sin^2\beta} m_h^2+\frac{\sin^2\alpha}{\sin^2\beta}m_H^2\right), \cr
\lambda_3 &=\frac{1}{v^2}\left(-M^2+2 m_{H^\pm}^2+\dfrac{\sin 2\alpha }{\sin 2\beta}(m_H^2-m_h^2)\right),\cr
\lambda_4 &= \frac{1}{v^2} \left(M^2+m_A^2-2 m_{H^\pm}^2\right) \cr
\lambda_5 &= \frac{1}{v^2} \left(M^2- m_A^2\right),
\end{align}
in an obvious notation in which we also replaced $M^2\equiv\dfrac{m_{12}^2}{\sin\beta \cos\beta}$. Conversely,
\begin{align}
m_H^2 &= M^2 \sin^2(\alpha-\beta)+\left(\lambda_1 \cos^2 \alpha \cos^2\beta+\lambda_2 \sin^2 \alpha \sin^2\beta+\frac{\lambda_{345}}{2}\sin 2\alpha \sin 2\beta\right)v^2,\cr
m_h^2 &= M^2 \cos^2(\alpha-\beta)+\left(\lambda_1 \sin^2 \alpha \cos^2\beta+\lambda_2 \cos^2 \alpha \sin^2\beta-\frac{\lambda_{345}}{2}\sin 2\alpha \sin 2\beta\right)v^2, \cr
m_A^2 &= M^2 - \lambda_5 v^2, \cr
m_{H^\pm}^2 &= M^2 - \frac{\lambda_{45}}{2} v^2.
\end{align}

To ensure that the scalar potential is bounded from below, the quartic parameters in Eq.~(\ref{v2hdm}) should satisfy~\cite{Gunion:2002zf}
\begin{equation}
\label{contrainte}
\lambda_{1,2}>0,\qquad{}\lambda_3>-(\lambda_1\lambda_2)^{1/2},\qquad{}\text{and}\qquad{}\lambda_3+\lambda_4-|\lambda_5|>-(\lambda_1 \lambda_2)^{1/2}.
\end{equation}
Stability of the vacuum ($ \partial V/\partial v_{1,2}=0$) amounts to solving
\bea\label{eq:STAB}
	m_{11}^2+\dfrac{\lambda_1 v_1^2}{2}+\dfrac{\lambda_3 v_2^2}{2}=\dfrac{v_2}{v_1} \left[m_{12}^2-(\lambda_4+\lambda_5)\dfrac{v_1 v_2}{2}\right],\cr 
	m_{22}^2+\dfrac{ \lambda_2v_2^2}{2}+\dfrac{\lambda_3 v_1^2}{2}=\dfrac{v_1}{v_2} \left[m_{12}^2-(\lambda_4+\lambda_5)\dfrac{v_1 v_2}{2}\right], 
\eea
which cannot be done analytically for $m_{12}\neq 0$. Instead, one can derive a condition that is necessary and sufficient for $V(\Phi_1,\Phi_2)$ to have a global minimum at $(v_1,v_2)$ and it reads~\cite{Barroso:2013awa},
\begin{equation}\label{eq:STAB2}
	m_{12}^2 \left(m_{11}^2-  m_{22}^2 \sqrt{\lambda_1/\lambda_2} \right)\left(\tan\beta-\sqrt[4]{\lambda_1/\lambda_2}\right)>0\,.
\end{equation}

Another generic constraint comes from the requirement of unitarity of the $S$-wave component of the partial wave decomposition of the full Higgs scattering amplitudes. That condition can be translated into a set of constraints on the quartic couplings in Eq.~(\ref{v2hdm}), which amounts to~\cite{unitarity}
\begin{equation}
\label{pertunit}
	|a_{\pm}|,|b_{\pm}|,|c_{\pm}|,|f_{\pm}|,|e_{1,2}|,|f_1|,|p_1|< 8 \pi,
\end{equation}
where 
\begin{align}
a_\pm &= \frac{3}{2}(\lambda_1+\lambda_2)\pm \sqrt{\frac{9}{4}(\lambda_1-\lambda_2)^2+(2\lambda_3+\lambda_4)^2}, \cr
b_\pm &= \frac{1}{2}(\lambda_1+\lambda_2)\pm \frac{1}{2}\sqrt{(\lambda_1-\lambda_2)^2+4\lambda_4^2}, \cr
c_\pm &= \frac{1}{2}(\lambda_1+\lambda_2)\pm \frac{1}{2}\sqrt{(\lambda_1-\lambda_2)^2+4\lambda_5^2}, \cr
e_1 &= \lambda_3 + 2\lambda_4 - 3 \lambda_5,  \qquad \qquad \qquad e_2 = \lambda_3- \lambda_5, \cr
f_+ &= \lambda_3 + 2\lambda_4 + 3 \lambda_5, \qquad \qquad \qquad  f_- = \lambda_3 + \lambda_5, \cr
f_1 &= \lambda_3 + \lambda_4, \qquad \qquad \qquad \qquad  \quad  p_1  = \lambda_3 - \lambda_4. 
\end{align}
 
We then generate random points in the parameter space by fixing $m_h=125.7(4)\, \rm{GeV}$, $v=246.22\, \rm{GeV}$, 
as well as $m_A=750(30)$~GeV, and by varying  
\bea
&&\tan\beta \in [1,35]\,,\qquad \quad\alpha \in [-\frac{\pi}{2},\frac{\pi}{2}]\,,\qquad m_H\in(m_h, 1.5 \, \mathrm{TeV}]\nn\\
&& m_{H^\pm}\in(m_W, 1.5 \, \mathrm{TeV}]\,,\quad  |M^2|={|m_{12}^2| \over \sin\beta\cos\beta} < (1.5\, \rm{TeV})^2 \,,
\eea 
to select those that are consistent with constraints given in Eqs.(\ref{contrainte}--\ref{pertunit}).

\begin{figure}[t!]
\centering
\includegraphics[width=0.5\linewidth]{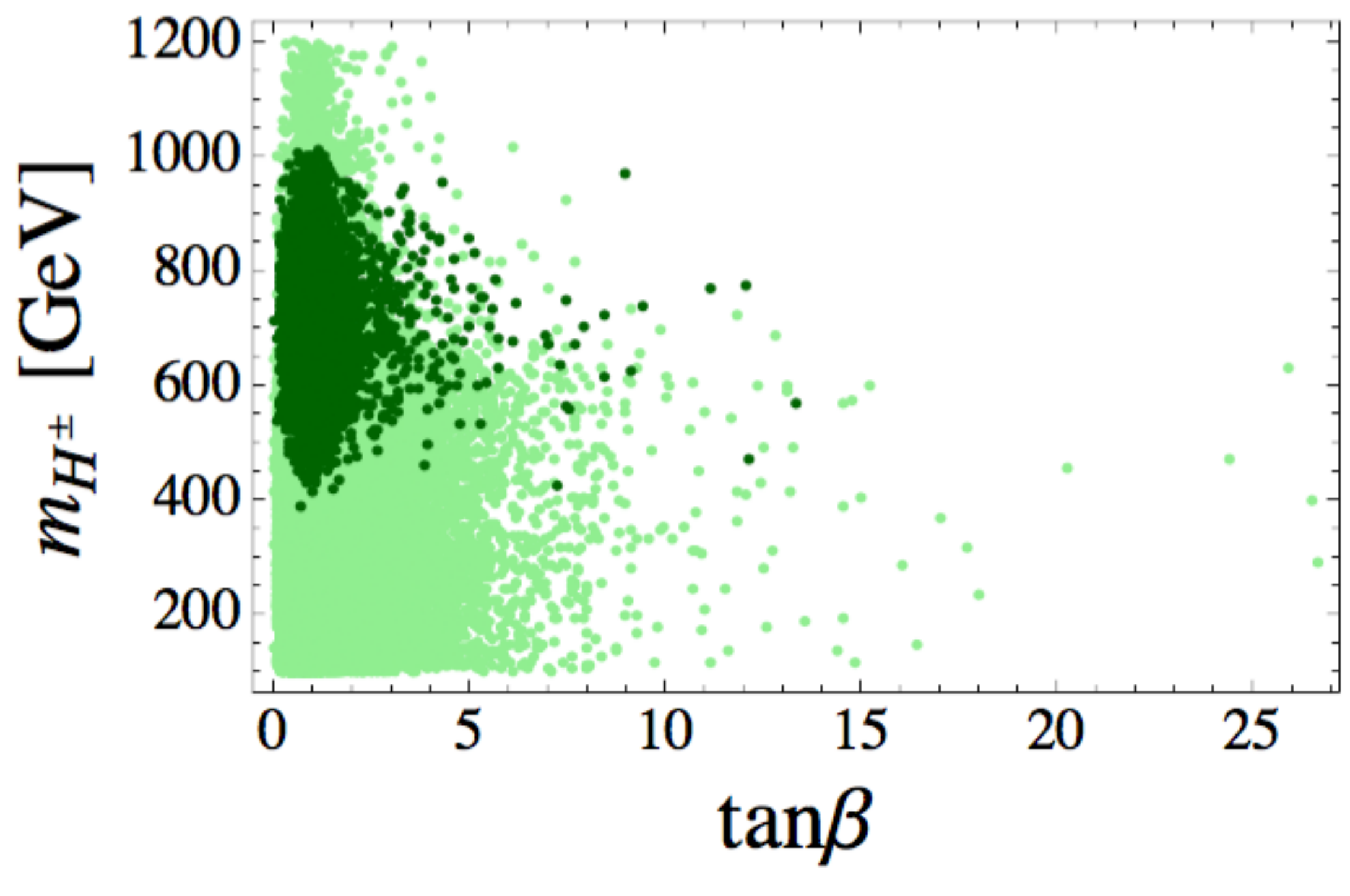}~\includegraphics[width=0.5\linewidth]{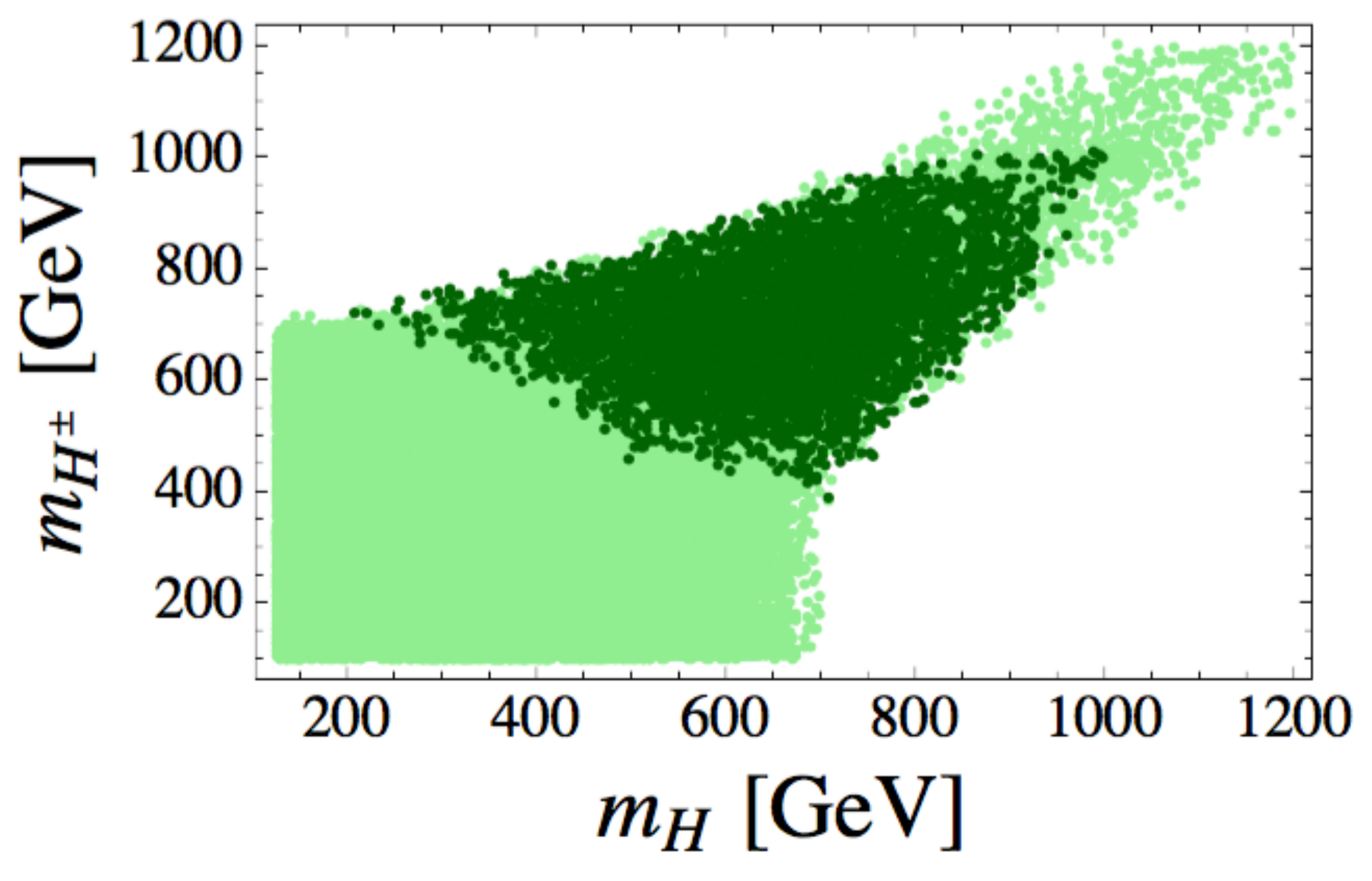}
\caption{\small\sl Result of the scan of the parameter space as indicated in the text and after imposing constraints given in Eqs.~(\ref{contrainte}--\ref{pertunit}). To better appreciate the effects of fixing $m_A=750(30)$~GeV, we also made the scan of parameters without fixing $m_A=750(30)$~GeV, shown in the plots by brighter points.}
\label{fig:1}
\end{figure}

Two interesting results of our scan are shown in Fig.~\ref{fig:1} where we see that the lower values of $\tan\beta$ are highly favored with most of the points being $\tan\beta \lesssim 5$, and that 
the mass of the other two Higgs states are bounded both from below and from above, namely,
\bea\label{eq:bounds}
400\ \gev \lesssim m_{H^\pm}\lesssim 1~{\rm TeV}, \qquad 200\ \gev \lesssim m_{H}\lesssim 1~{\rm TeV}.
\eea
Furthermore we observe that the $Z_2$-symmetry breaking term cannot be
excessively large and it reproduces $m_A=750(30)$~GeV, for $|M|\in
(200,800)$~GeV. Finally the resulting points are concentrated in the region of
$|\cos(\beta - \alpha )| \lesssim 0.3$, which then means that the couplings
$g_{hVV}> g_{HVV}$ for $V$ being either $W$ or $Z$.~\footnote{\label{footnote:couplings} We remind the
  reader that $g_{HVV} = 2 \cos(\beta - \alpha ) m_V^2/v$ and $g_{hVV} = 2
  \sin(\beta - \alpha ) m_V^2/v$.} This result agrees with the findings of
ref.~\cite{Corbett:2015ksa}.

Another interesting feature is that one cannot impose the degeneracy $m_h=m_H$
and scan over the parameter space as indicated above (but without fixing
$m_A$). While most of $\tan\beta$ points remain small, the values of
$\cos(\alpha-\beta)$ are equidistributed between $-1$ and $1$, but one then gets an upper
bound on $m_A \lesssim 700$~GeV, inconsistent with the state supposedly
observed at LHC. We should also mention that the direct experimental searches of the non-Standard Model Higgs states also restrict 
$\cos(\alpha-\beta)$ to small values, as recently discussed in Refs.~\cite{lydia,others}.

\section{Including fermions}

As it is well known, in order to avoid the tree level flavor changing neutral currents (FCNC), one imposes that 
the fermions of definite charge and chirality couple to a specific Higgs doublet. In this way one distinguishes various types of 2HDM: In Type I models all fermions couple to the same Higgs doublet ($\Phi_2$, by convention); In Type II the right-handed (RH) quarks with charge ${\cal Q}=2/3$ couple to the doublet $\Phi_2$, whereas those with ${\cal Q}=-1/3$ to the doublet $\Phi_1$. A slight modification of the latter rule leads to two other types: Type X  2HDM (or {\sl lepton-specific}), and type Z (or {\sl flipped}) one. Their coupling patterns to quarks and leptons are listed in Tab.~\ref{THDM}. 
More general, in terms of Yukawa couplings, is the so called Type III 2HDM in which the couplings to fermions are all to be fixed by the data~\cite{Crivellin:2013wna} which is often impractical because of too many free parameters so that one has to resort to additional assumptions such as minimal flavor violation (MFV)~\cite{mfv,CONSTRAINTS}, the natural flavor conservation~\cite{Buras:2010mh}, or the aligned 2HDM~\cite{pich} where the minimal flavor violation is ensured by assuming proportionality between the matrices of Yukawa couplings to the two Higgs doublets.

\begin{table}[h]
\renewcommand{\arraystretch}{1.5}
\begin{center}
\begin{tabular}{cccc}    \hline
\hspace*{4mm}{Model}\hspace*{6mm}  &  $u_R$ & $d_R$ & $\ell_R$\hspace*{4mm} \\ \hline
Type I     & $\Phi_2$ & $\Phi_2$ & $\Phi_2$ \\ 
Type II    & $\Phi_2$ & $\Phi_1$ & $\Phi_1$\\
Type $X$    & $\Phi_2$ & $\Phi_2$ & $\Phi_1$\\
Type $Z$    & $\Phi_2$ & $\Phi_1$ & $\Phi_2$\\ 
 \hline
\end{tabular}
\caption{\small \sl Flavor conserving models and the respective Yukawa couplings of the quarks $u_R$ (charge ${\cal Q}=2/3$), $d_R$ (charge ${\cal Q}=-1/3$) and leptons $\ell_R$ with the Higgs doublets.}
\label{THDM}
\end{center}
\end{table}

The Yukawa interaction Lagrangian for the neutral currents can be written as~\cite{Branco:2011iw},
\begin{align}
\label{lagrangianYukawa}
\mathcal{L}_\mathrm{Y}^{nc} &= -\sum_{f={u,d,\ell}} \frac{m_f}{v}\left( C_{hf} \ \bar{f} f h+ C_{Hf} \ \bar{f} f H	-i C_{Af}\ \bar{f}\gamma_5 f A\right), 
\end{align}
where the sum runs over up- and down-type quarks, as well as leptons. Here we focus on the coupling $C_{Af}$ which depends on $\tan\beta$ and is given for various types of 2HDM in Tab.~\ref{nccouplings}. 
\begin{table}[h]
\renewcommand{\arraystretch}{1.5}
\begin{center}
\begin{tabular}{ccccc}   \hline
\hspace*{4mm}Model &  Type I & Type II & Type $X$ & Type $Z$ \hspace*{4mm}\\ \hline
$C_{Au}$    & $\cot\beta$ & $\cot\beta$ & $\cot\beta$ & $\cot\beta$ \\
$C_{Ad}$  & $-\cot\beta$ & $\tan\beta$ & $-\cot\beta$ & $\tan\beta$\\
$C_{A\ell}$   & $-\cot\beta$ & $\tan\beta$ & $\tan\beta$ & $-\cot\beta$ \\
 \hline
\end{tabular}
\caption{\small\sl Couplings appearing in the lagrangian~(\ref{lagrangianYukawa}) for the models of type I, II, Lepton-specific ($X$) and Flipped ($Z$) \cite{Branco:2011iw}.}
\label{nccouplings}
\end{center}
\end{table}
\begin{figure}[t!]
\centering
\includegraphics[width=0.5\linewidth]{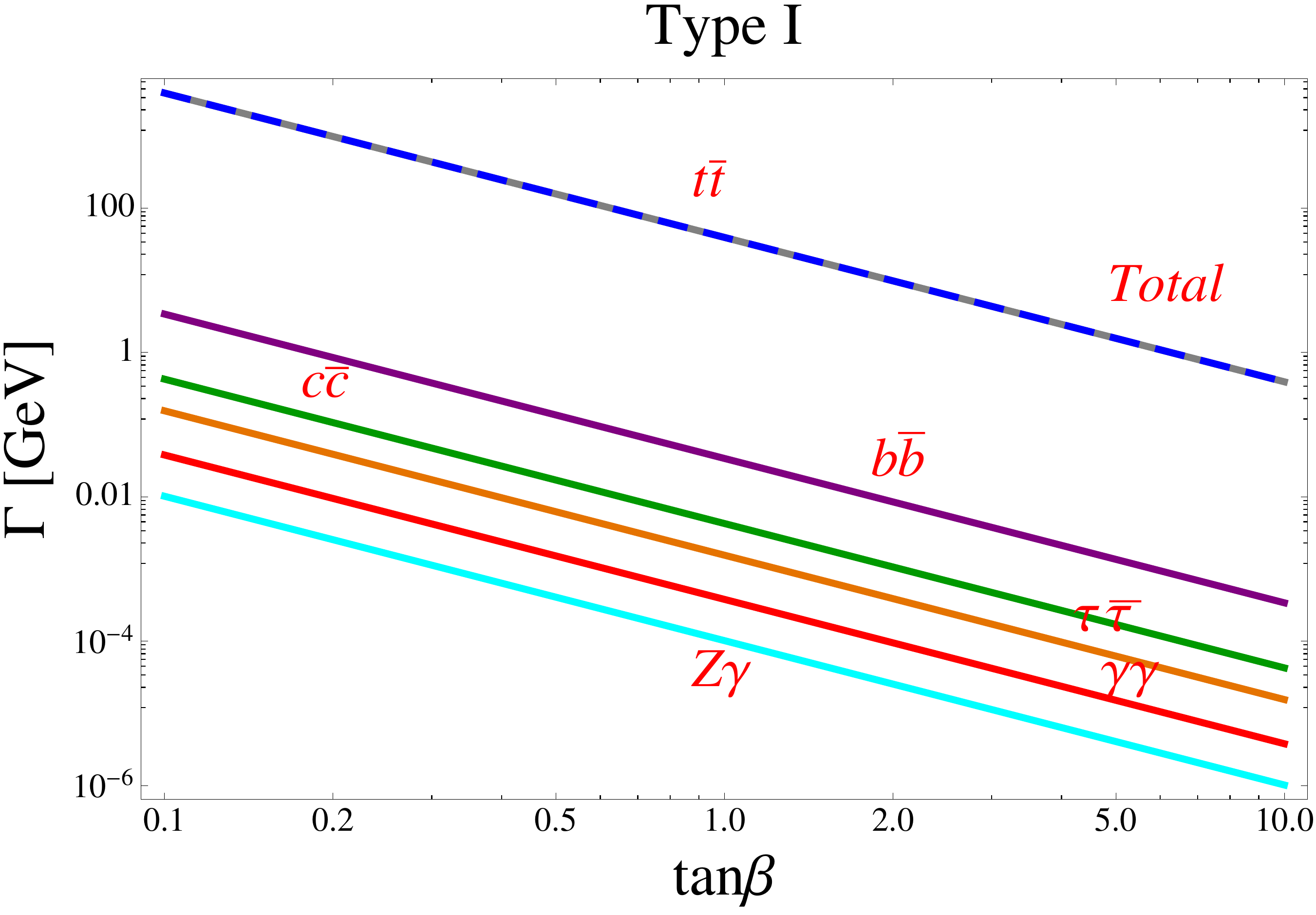}~\includegraphics[width=0.5\linewidth]{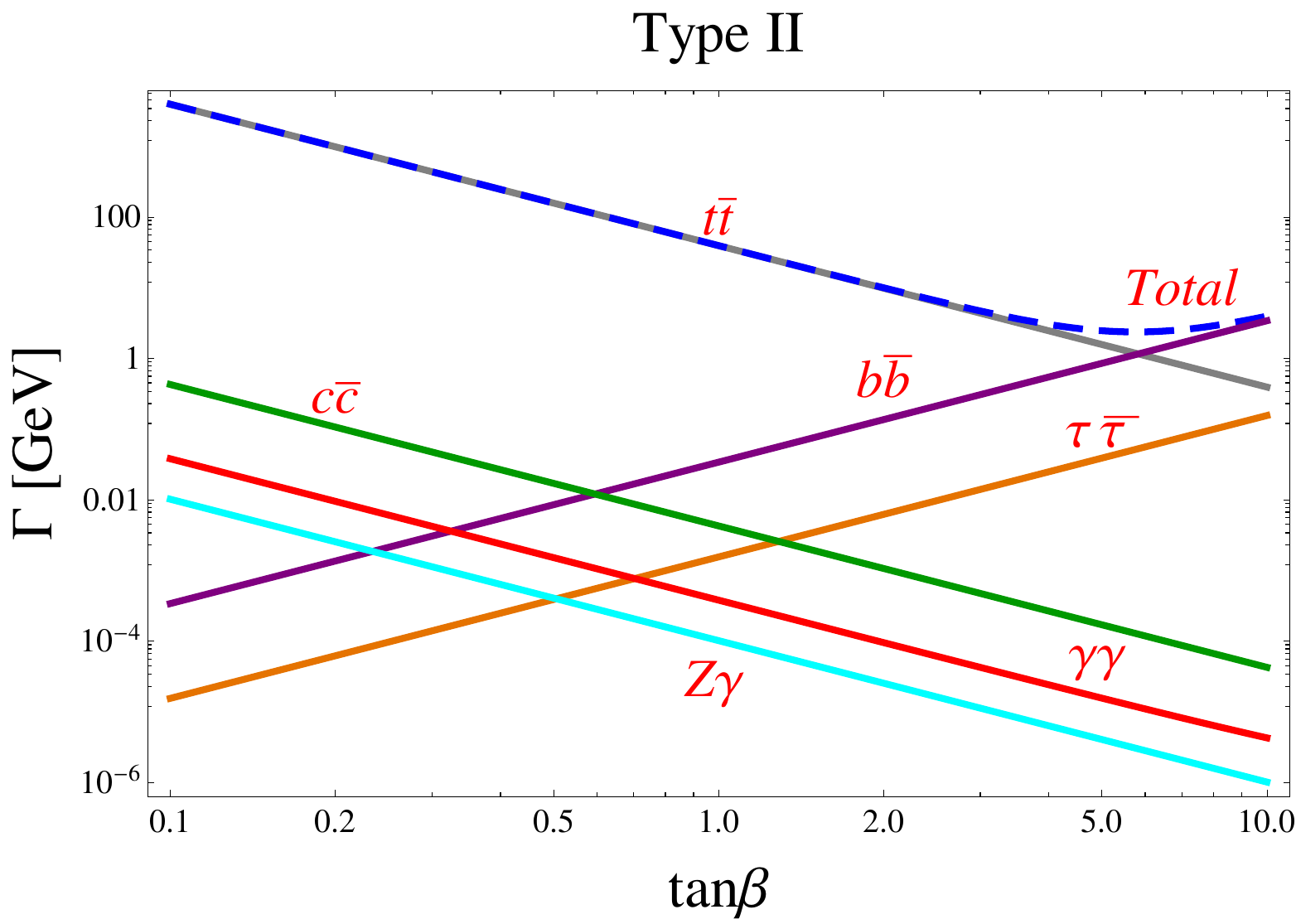}\\
\includegraphics[width=0.5\linewidth]{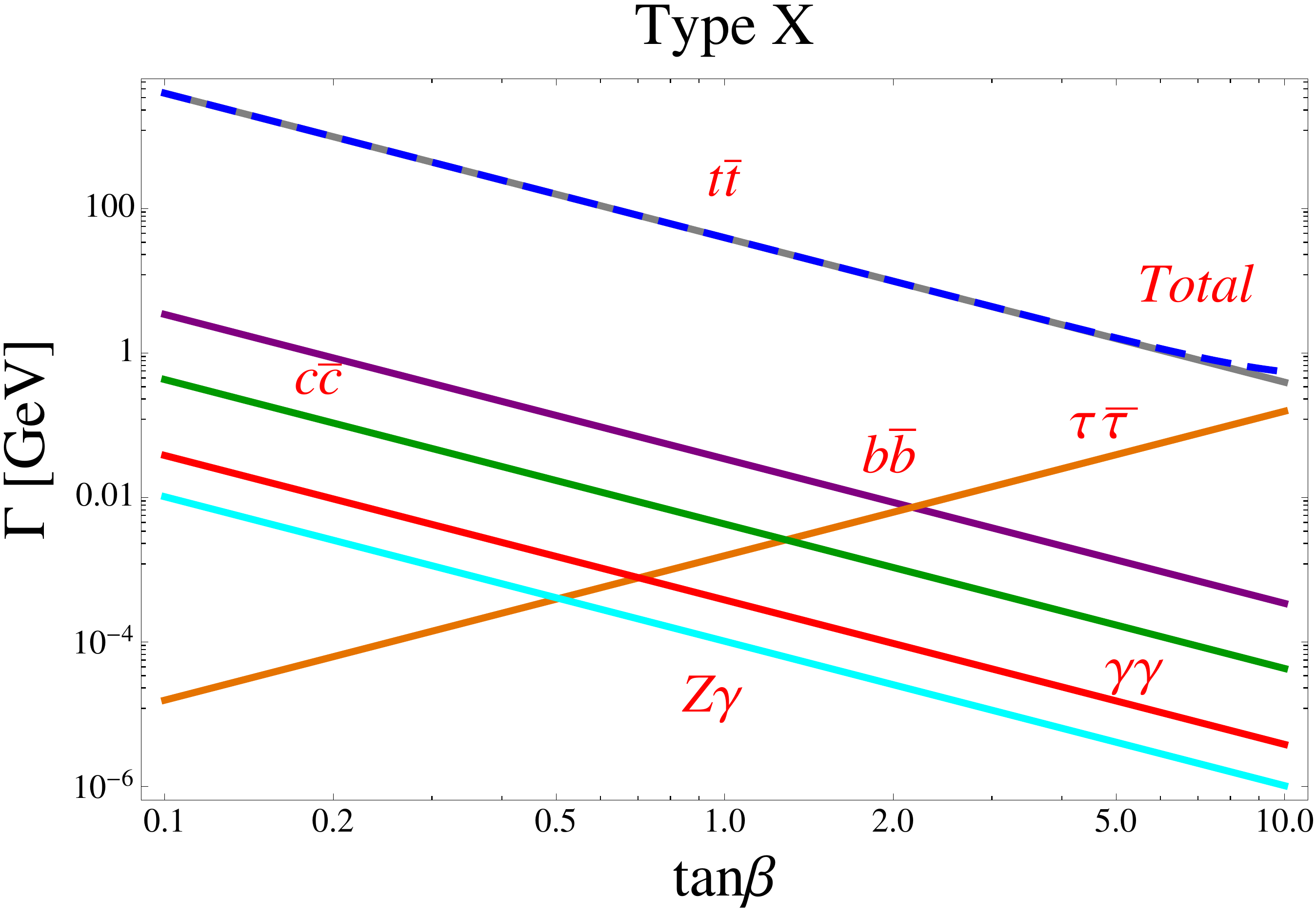}~\includegraphics[width=0.5\linewidth]{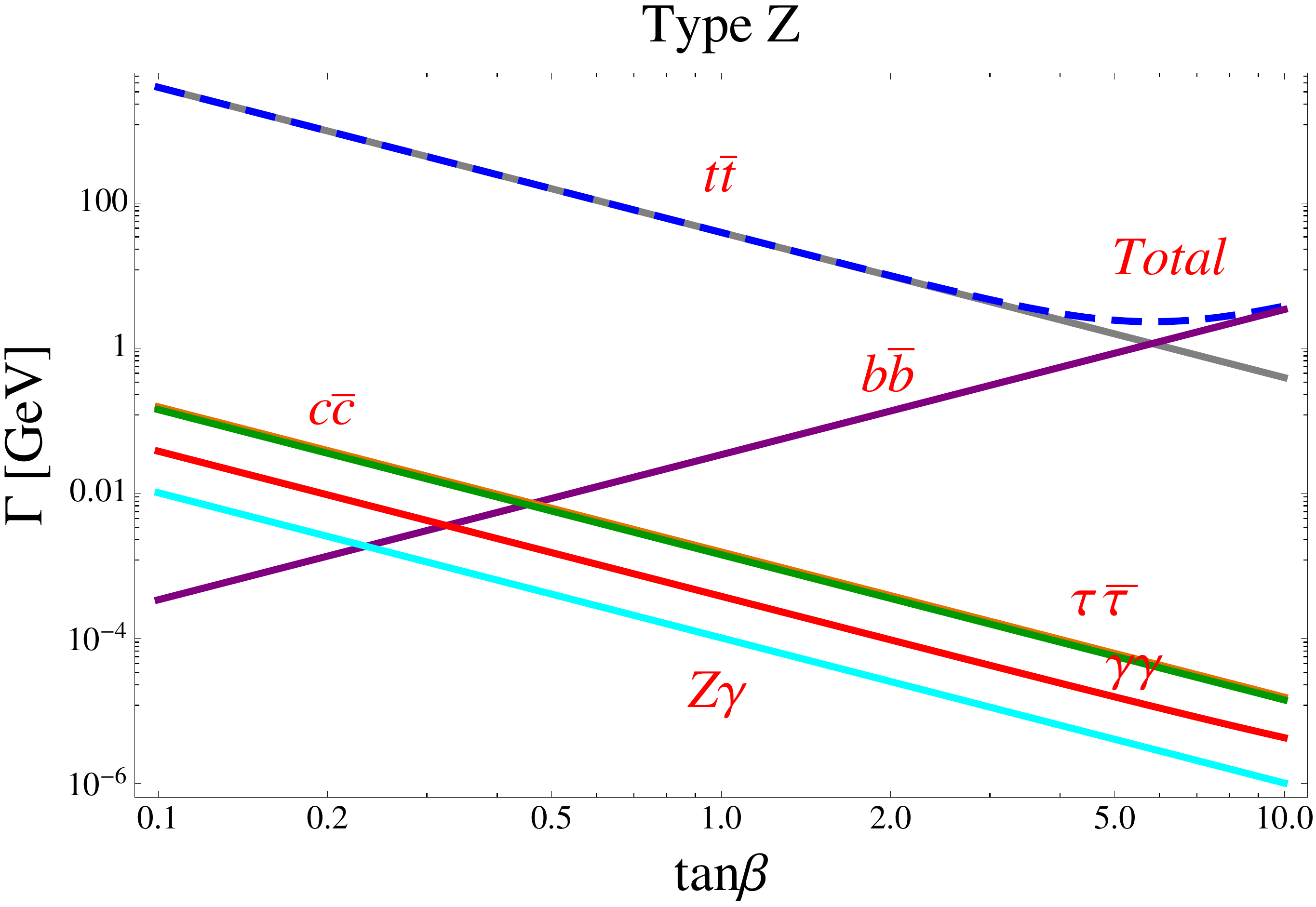}
\caption{\small \sl Partial decay widths as functions of $\tan
\beta$ for all four types of 2HDM considered in this paper. Full decay width is depicted by the dashed curve and it is most often indistinguishable from $\Gamma(A\to t \bar t)$. 
}
\label{fig:2}
\end{figure}
With this in mind, the expression for the decay width of the CP-odd Higgs to two fermions reads,
\bea
\Gamma(A\to f\bar f)= {N_c\over 8\pi} |C_{Af}|^2 {m_f^2 \over v^2} \sqrt{m_A^2-4 m_f^2}\,
\eea 
where $N_c=3$ for quarks, and $1$ otherwise.  Quite obviously the full width $\Gamma(A)$ will be highly dominated by the top quark. 
As for the decay to two photons, one has~\cite{Spira:1995rr},
\bea
\Gamma(A\to \gamma\gamma)= {\alpha^2m_A^3\over 64 v^2\pi^3}\left| \sum_f N_c Q_f^2 C_{Af} F\left({m_A^2\over 4m_f^2}\right)\right|^2,
\eea
where the triangle loop induces the factor, 
\bea
F(x)=\left\{
\begin{array}{ll}  \displaystyle
\frac{1}{x}\arcsin^2\sqrt{x} & x\leq 1 \\
\displaystyle -\frac{1}{4x}\left[ \log\frac{1+\sqrt{1-x^{-1}}}
{1-\sqrt{1-x^{-1}}}-i\pi \right]^2 \hspace{0.5cm} & x>1\,.
\end{array} \right.
\eea
\begin{table}[ht!]
\renewcommand{\arraystretch}{1.5}
\centering{}%
\resizebox{\textwidth}{!}{
\tabcolsep=0.09cm
\begin{tabular}{|c|cccc|}
\hline
\hspace*{2.9cm} & Type I & Type II & Type $X$  & Type $Z$ \\ \hline\hline
\hspace*{-1.2cm}\underline{$\;\tan\beta=1\;$} &   &  &  &  \\ 
 $\Gamma(A) [\gev]$ & $39.3$ & $39.3$ & $39.3$  & $39.3$ \\ 
 $B(A\to \tau\tau)$ & $4.0  \times 10^{-5}$ &  $4.0  \times 10^{-5}$ & $4.0  \times 10^{-5}$  & $4.0  \times 10^{-5}$ \\ 
 $B(H^-\to \tau\bar{\nu})$ & $(1.1-1.6)  \times 10^{-4}$ &  $(1.1-1.6)  \times 10^{-4}$ & $(1.1-1.6)  \times 10^{-4}$  & $(1.1-1.6)  \times 10^{-4}$ \\ \hline
\hspace*{-1.2cm}\underline{$\;\tan\beta=2\;$} &   &  &  &  \\ 
 $\Gamma(A) [\gev]$ & $9.8$ & $10.0$ & $9.8$  & $9.9$ \\ 
 $B(A\to \tau\tau)$ & $4.0  \times 10^{-5}$ & $6.2  \times 10^{-4}$ & $6.2 \times 10^{-4}$  & $4.0  \times 10^{-5}$ \\ 
$B(H^-\to \tau\bar{\nu})$  & $(1.1-1.6)  \times 10^{-4}$ & $(1.8-2.5)  \times 10^{-3}$ & $(1.8-2.5) \times 10^{-3}$  & $(1.1-1.6)  \times 10^{-4}$ \\ \hline
\hspace*{-1.2cm}\underline{$\;\tan\beta=5\;$} &   &  &  &  \\ 
 $\Gamma(A) [\gev]$ & $1.6$ & $2.5$ & $1.6$  & $2.4$ \\ 
 $B(A\to \tau\tau)$ & $4.0  \times 10^{-5}$ & $1.6  \times 10^{-2}$ & $1.6 \times 10^{-2}$  & $2.6  \times 10^{-5}$ \\ 
$B(H^-\to \tau\bar{\nu})$  & $(1.1-1.6) \times 10^{-4}$ & $(4.5-6.4)  \times 10^{-2}$ & $(6.5-9.1) \times 10^{-2}$  & $(7.6-11)  \times 10^{-5}$ \\ \hline
\hspace*{-1.0cm}\underline{$\;\tan\beta=10\;$} &   &  &  &  \\ 
 $\Gamma(A) [\gev]$ & $0.4$ & $4.0$ & $0.6$  & $3.8$ \\ 
 $B(A\to \tau\tau)$ & $4.0  \times 10^{-5}$ & $3.9  \times 10^{-2}$ & $3.9 \times 10^{-2}$  & $4.1  \times 10^{-6}$ \\ 
$B(H^-\to \tau\bar{\nu})$  & $(1.1-1.6) \times 10^{-4}$ & $(1.1-1.6)  \times 10^{-1}$ & $(5.3-6.2) \times 10^{-1}$  & $(1.3-1.9)  \times 10^{-5}$ \\ \hline
\end{tabular}
}
\caption{\footnotesize{}\label{tab:2} \sl Results for the decay width of the CP-odd Higgs boson of mass $m_A=750(30)$~GeV, for four different values of $\tan\beta$ discussed in the previous section, and for four different types of 2HDM. 
Furthermore we give the branching fraction of the $A\to \tau\tau$ and $H^-\to \tau\bar{\nu}$ decay modes. The value of $m_{H^-}$ is varied within the bounds quoted in Eq.~(\ref{eq:bounds}).}
\end{table}
The expression for the decay width of $A\to Z\gamma$ reads~\cite{Gunion:1991cw},
\bea
\Gamma(A\to Z\gamma)= {\alpha^2m_A^3 N_c\over 384 v^2 \pi^3} C_{At}^2  \left({1- (8/3) \sin^2\theta_w \over \sin\theta_w\cos\theta_w}\right)^2 \left( 1- \frac{m_Z^2}{m_A^2}\right)^3,
\eea
and its contribution to the full decay width is smaller than the other modes discussed above. 

In Fig.~\ref{fig:2} we plot the partial decay width for $A\to t\bar t$, $A\to b\bar b$,  $A\to \tau\tau$, $A\to \gamma\gamma$ and $A\to Z\gamma$ for all four types of the 2HDM discussed here. 
As expected, the $A\to t\bar t$ mode is largely dominant and essentially saturates the width of the CP-odd Higgs boson, $\Gamma(A)$, 
the results of which are given in Tab.~\ref{tab:2} for four different values of $\tan \beta$. Since the experimenters are searching for the $A\to \tau\tau$ mode, in the same table we also give the values of 
$B(A\to \tau\tau)$ for all four types of 2HDM considered here. It is worth emphasizing that for small values of $\tan\beta$ the width $\Gamma(A)$ is quite large and can accommodate the observation made by ATLAS, namely that the better fits are obtained 
for the new resonance having a width $\sim 40$~GeV. Similar conclusion holds true for the $H$ boson, the coupling of which to $t\bar t$ is proportional to $1/\tan\beta$ for small $\cos(\beta -\alpha)$.

Finally, with the above ingredients we can also compute the dominant decay channel $H^\pm \to tb$, the decay width of which is given by, 
\bea
\Gamma(H^\pm \to tb ) = {1\over 8 \pi} {|V_{tb}|^2 \over v^2}  \left( C_{Ab}^2 m_b^2  + C_{At}^2 m_t^2  - 4 C_{At}C_{Ab} {m_b^2m_t^2\over m_{H^\pm}^2-m_t^2} \right) {(m_{H^\pm}^2-m_t^2)^2\over m_{H^\pm}^3},
\eea
while for the leptonic decay we have
\bea
\Gamma(H^\pm \to \ell \bar \nu_\ell ) = {m_{H^\pm}\over 8 \pi} \left( {C_{A\ell} m_\ell\over v}\right)^2  .
\eea
The results for the branching fraction $B(H^\pm \to \tau \nu_\tau)$ are given in Tab.~\ref{tab:2}. 
Before concluding this Section we would like to emphasise once again that the bounds on the Higgs states are derived by considering the general theoretical arguments. 
The most significant bound comes from the tree-level unitarity constraints~(\ref{pertunit}), and the effects of heavy fermions (beyond the Standard Model) would enter only through loops and are therefore unlikely to significantly alter the statements we made about the Higgs states. 

\section{Electroweak Precision Tests\label{sec:EW}}
\begin{figure}[h!]
\centering
\hspace*{-.4cm}\includegraphics[width=0.61\linewidth]{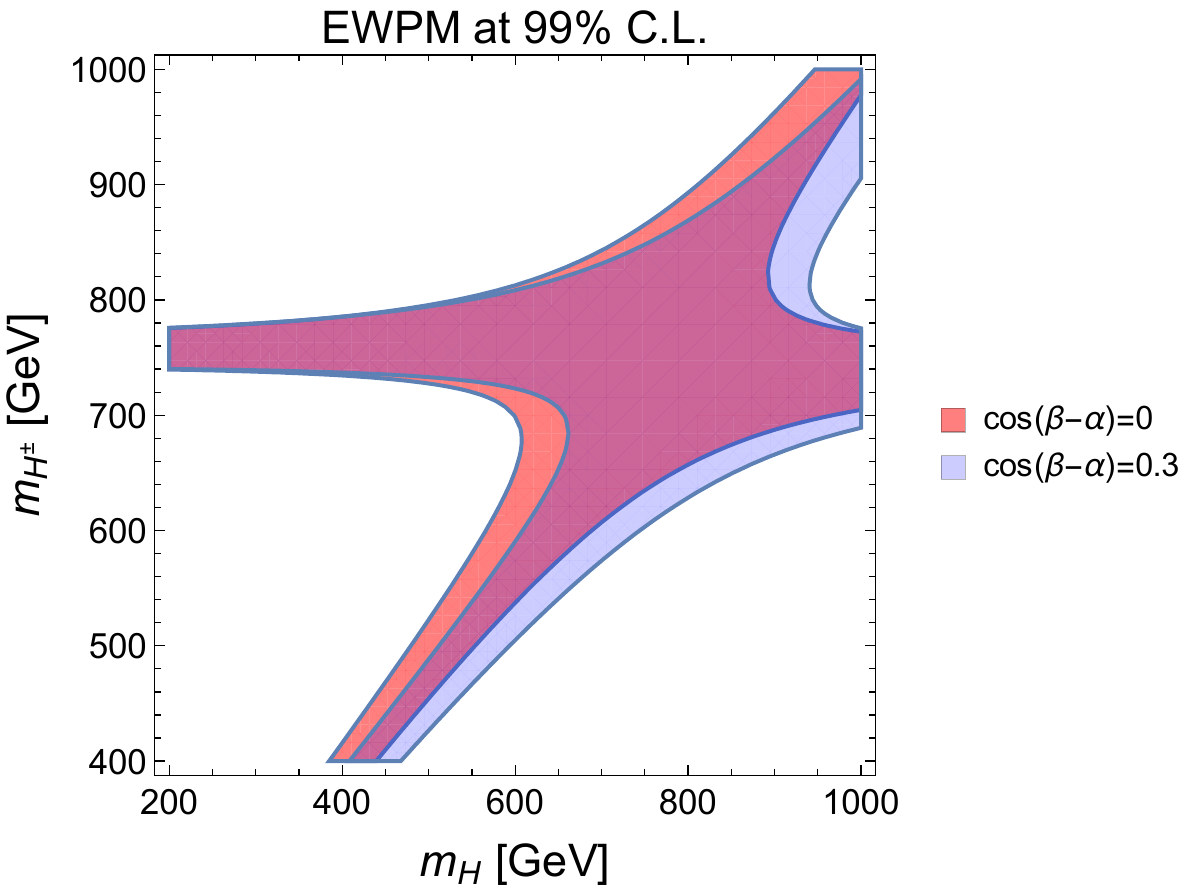} 
\caption{\small \sl Allowed region by the $S$, $T$ and $U$ parameters at $99\%$ CL.
}
\label{fig:4}
\end{figure}
As it is well known, the additional scalar states present in a 2HDM contribute to the gauge bosons vacuum polarizations, and are as such constrained by electroweak precision data. The scalar contributions to the Peskin-Takeuchi parameters $S$, $T$, and $U$ for the 2HDM case can be found e.g. in ref.~\cite{Barbieri:2006bg}. In order to compute the related bounds on the spectrum, we used the latest Gfitter values for the
best fit, uncertainties and covariance matrix~\cite{Baak:2014ora},
\begin{equation}
  \begin{aligned}
    & \Delta S^{SM} = 0.05\pm0.11,\\
    & \Delta T^{SM} = 0.09\pm0.13,\\
    & \Delta U^{SM} = 0.01\pm0.11,\\
  \end{aligned}
\qquad\qquad
V = \left(\begin{array}{ccc}
1 & 0.90 & -0.59\\
0.90 & 1 & -0.83\\
-0.59 & -0.83 & 1
\end{array}\right),
\end{equation}
composing the $\chi^2$ function as
\begin{equation}
  \chi^2= \sum_{i,j}(X_i - X_i^{\rm SM})(\sigma^2)_{ij}^{-1}(X_j - X_j^{\rm SM}),
\end{equation}
with $X_i=\Delta S, \Delta T, \Delta U$ and the covariance matrix
$\sigma^2_{ij}\equiv\sigma_iV_{ij}\sigma_j$, in which
$(\sigma_1,\sigma_2,\sigma_3)=(0.11,0.13,0.11)$. 

In Fig.~\ref{fig:4} we show the region allowed at 99\% CL by electroweak precision data 
in the plane $m_H$ versus $m_{H^\pm}$. Since the coupling between the additional scalars and the gauge bosons depend on $\cos(\beta-\alpha)$ (see footnote~\ref{footnote:couplings}), we present two representative cases: $\cos(\beta-\alpha)=0$ and $\cos(\beta-\alpha)=0.3,$ which according to Sec.~\ref{sec:2}, are the minimum and maximum value allowed by our scan. Note that most of the points which were previously allowed, Fig.~\ref{fig:1}, are still not excluded. Let us stress that, as already pointed out in the introduction, additional states which may be needed to increase the production cross section $\sigma(gg \to A)$ may affect significantly Fig.~\ref{fig:4}, but in a model dependent way. The analysis of the electroweak precision measurements in these extended models must be done case by case, and is beyond the scope of the paper.

\section{Low energy physics observables\label{sec:flavor}}

Since the charged Higgs boson is now fully bounded [cf.~Eq.(\ref{eq:bounds})], the contribution from the charged Higgs can modify the low energy 
decay rates of the leptonic and semileptonic processes which generally agree with the Standard Model predictions within the error bars. To that end we add a term 
involving the couplings to the scalar sector to the effective Hamiltonian of the Standard Model at low energies, namely 
\begin{align}
	\label{heff}
	\mathcal{H}_{\text{eff}}=\sqrt{2} G_F V_{ud} &\Big{[}  (\overline{u}\gamma_\mu d)(\overline{\ell}_L \gamma^\mu \nu_L)+g_S(\mu)(\overline{u}d)(\overline{\ell}_R\nu_L)  +g_P(\mu)(\overline{u}\gamma^5 d)(\overline{\ell}_R\gamma^5 \nu_L) \Big{]} +\text{h.c.}  ,
\end{align}
where $u$ and $d$ stand for the generic up- and down-type quark flavor. Using this Hamiltonian one can easily compute the semileptonic and the leptonic decay rates for the specific channels, e.g. $B\to D\tau \nu_\tau$ and $B\to \tau \nu_\tau$, and we obtain
\begin{align}\label{eq:LSL}
&{dB\over dq^2} (B\to D \tau \nu_\tau)  =   {\tau_B G_F^2 \over 192\pi^3 m_B^3} |V_{cb}|^2 |f_+(q^2)|^2  \left[ c_+(q^2) + c_0(q^2)  \left( 1+g_S {q^2\over m_\tau (m_b-m_c)}\right)^2 \left| {f_0(q^2)\over f_+(q^2)}\right|^2 \right],\cr
&B(B\to \tau \nu_\tau) =\tau_B  {G_F^2 m_B m_\tau^2\over 8\pi} |V_{ub}|^2  f_B^2 \left( 1 - \frac{m_\tau^2}{m_B^2} \right)^2 \left( 1 - g_P {m_B^2\over m_\tau  m_b } \right)^2,
\end{align}
with $\tau_B$ being the $B$-meson lifetime, $m_\tau^2 \leq q^2 \leq (m_B-m_D)^2$,  
\begin{align}
&c_+(q^2) =  \lambda^{3/2}(q^2) q^2 \left[ 1-\frac{3}{2}{m_\tau^2 \over q^2} +\frac{1}{2} \left(  {m_\tau^2 \over q^2}\right)^3 \right], \cr 
&c_0(q^2) =  \lambda^{1/2}(q^2) m_\tau^2 \frac{3}{2}{m_B^4 \over q^2}
\left( 1-{m_\tau^2 \over q^2}\right)^2  \left( 1-{m_D^2 \over m_B^2} \right)^2,
\end{align}
and $\lambda(q^2) = [q^2 - (m_B+m_D)^2][q^2 - (m_B-m_D)^2]$. The decay constant ($f_B$) and the form factors [$f_{+,0}(q^2)$] are defined via,
\bea
&&\langle 0\vert \bar u \gamma_\mu \gamma_5 b\vert B(p)\rangle = i f_B p_\mu\,, \cr
&&\langle D(p')\vert \bar c \gamma_\mu  b\vert B(p)\rangle =  \left(p_\mu + p'_\mu - {m_B^2-m_D^2\over q^2}q_\mu \right) f_+(q^2) + {m_B^2-m_D^2 \over q^2} q_\mu  f_0(q^2)   \,. 
\eea
Notice that we consider the pseudoscalar to pseudoscalar meson decay for which the decay form factors are better controlled through numerical simulations of QCD on the lattice~\cite{FLAG}. 
As it can be seen from the above expressions, for $g_{S,P}\neq 0$ the helicity suppression is lifted and the contribution coming from coupling to the charged scalar could be important.  
The explicit expressions for $g_{S,P}$, in terms of the quark and lepton masses as well as $m_{H^\pm}$ and $\tan\beta$, in various types of 2HDM read:

\begin{center}\renewcommand{\arraystretch}{1.8}
\begin{tabular}[H]{ccc}   
 \hline
Type & $g_S$ & $g_P$ \\ \hline
I  & $- \dfrac{ m_\ell }{{m^2_{H^\pm}}} (m_d-m_u) \cot^2\beta $ & $ \dfrac{ m_\ell }{{m^2_{H^\pm}}} (m_d+m_u) \cot^2\beta $ \\
II  & $- \dfrac{ m_\ell }{{m^2_{H^\pm}}} (m_u + m_d \tan^2\beta )$ & $- \dfrac{ m_\ell }{{m^2_{H^\pm}}} (m_u - m_d \tan^2\beta )$ \\
$X$  & $ \dfrac{ m_\ell }{{m^2_{H^\pm}}} (m_d - m_u )$ & $- \dfrac{ m_\tau }{{m^2_{H^\pm}}} (m_d + m_u )$ \\
$Z$  & $ \dfrac{ m_\ell }{{m^2_{H^\pm}}} (m_d + m_u \cot^2\beta )$ & $- \dfrac{ m_\ell }{{m^2_{H^\pm}}} (m_d - m_u \cot^2\beta )$ \\
 \hline
\end{tabular}
\end{center}

\begin{figure}[t!]
\centering
\hspace*{-.4cm}\includegraphics[width=0.51\linewidth]{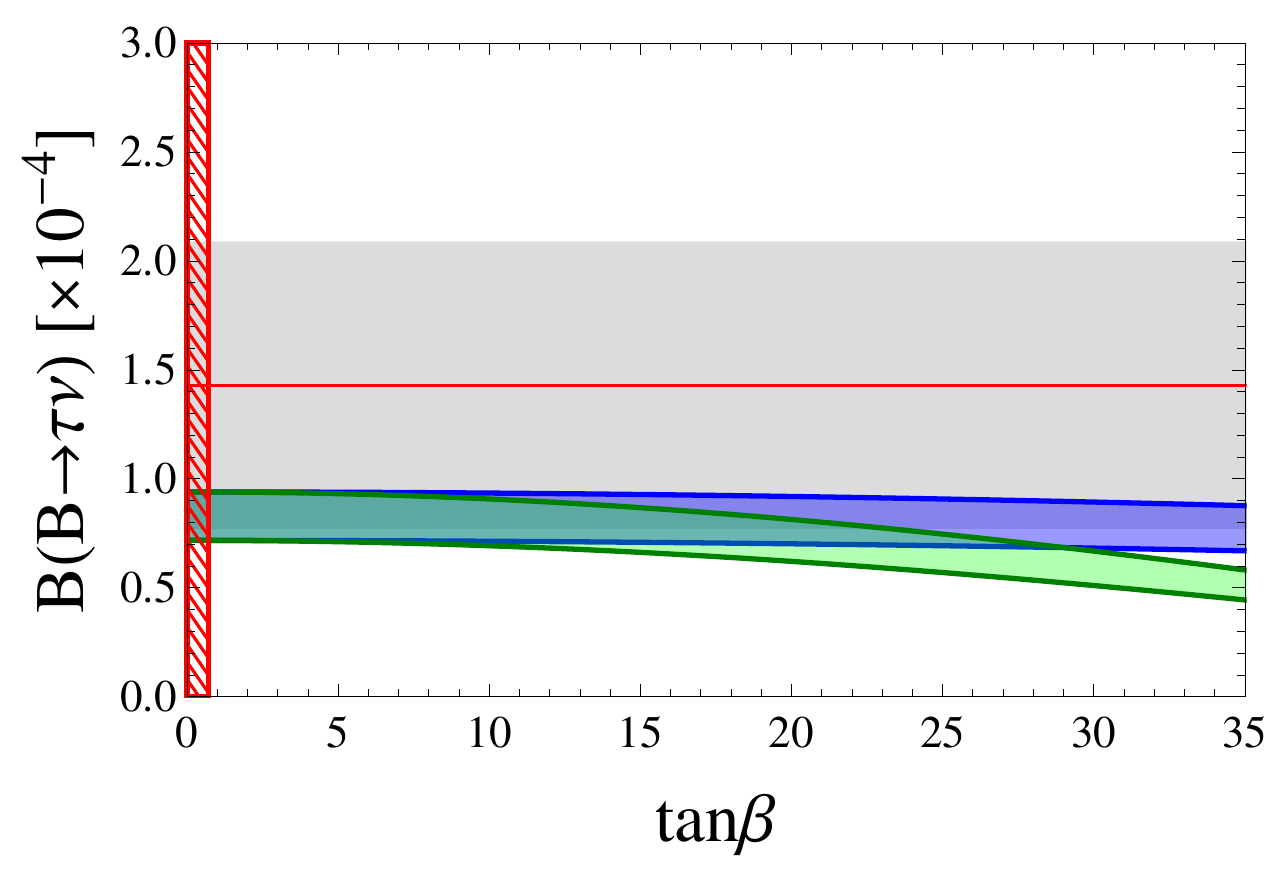}~\includegraphics[width=0.52\linewidth]{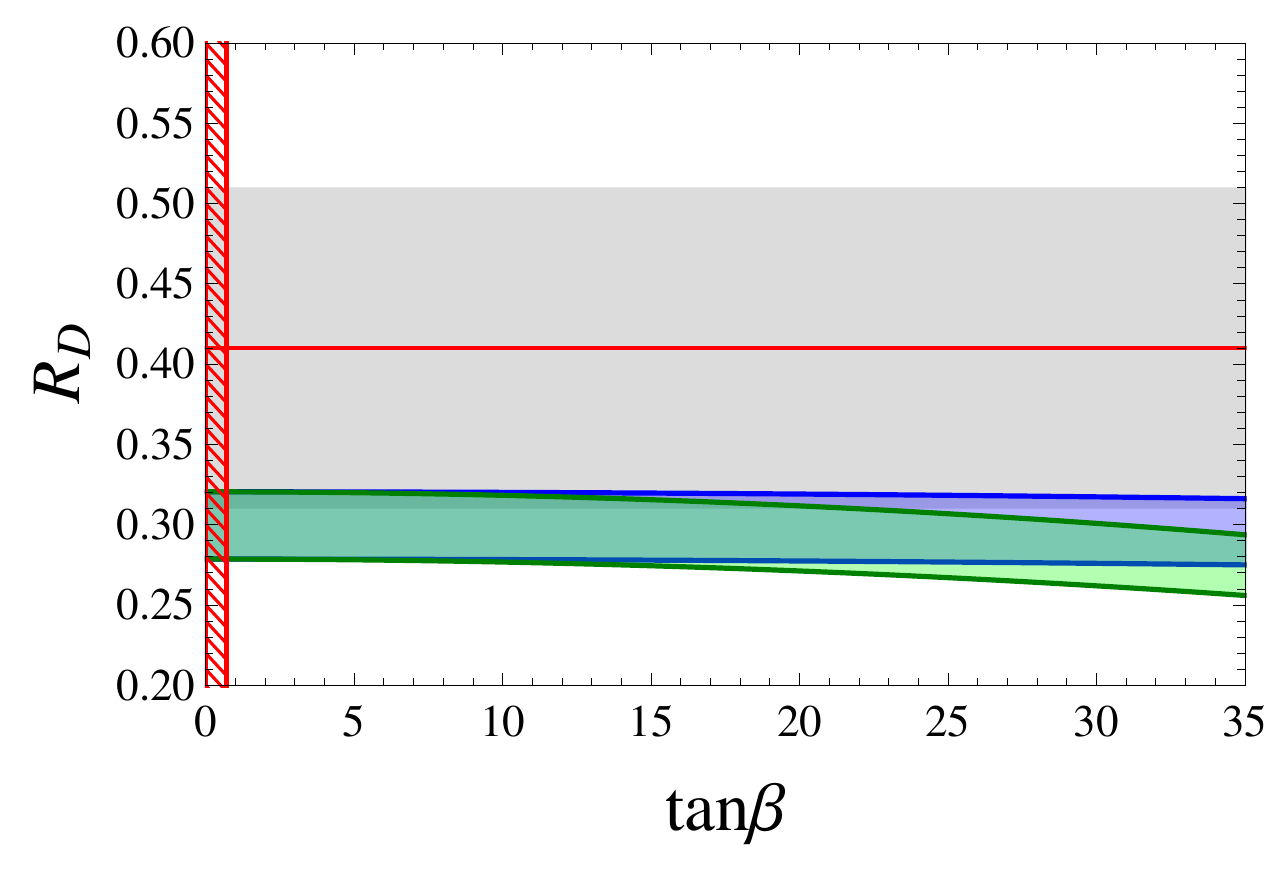}  
\caption{\small \sl  $B(B\to \tau \nu_\tau)$ and $R_D$ are computed in the Type~II 2HDM using $m_{H^\pm}=400$~GeV and $m_{H^\pm}=1$~TeV, and it is compared to the experimental values (gray bands) at $2\sigma$-level. Central experimental values are depicted by the full horizontal lines. 
The band showing deviation from the Standard Model for large values of $\tan\beta$, in both plots, corresponds to $m_{H^\pm}=400$~GeV. The hatched stripes, instead, correspond to low $\tan\beta \lesssim 0.7$, excluded by $B(B_s\to \mu^+\mu^-)$.}
\label{fig:3}
\end{figure}

By averaging the values obtained by BaBar~\cite{Lees:2012ju} and Belle~\cite{Kronenbitter:2015kls}, we have $B(B\to \tau \nu_\tau)=1.44(32) \times 10^{-4}$, which we then combine with $f_B =188(6)$~MeV~\cite{FLAG}, 
and $V_{ub}=3.6(1)\times 10^{-3}$ as obtained from the global fit by UTfit and CKM-fitter~\cite{CKMology}, to conclude that for all types of 2HDM considered here the resulting value for $B(B\to \tau \nu_\tau)_{\rm 2HDM}$ is consistent with experiment and is 
practically indistinguishable from the Standard Model predictions. Only for large values of $\tan\beta \gtrsim 20$ the $B(B\to \tau \nu_\tau)_{\rm 2HDM}$ may differ considerably from $B(B\to \tau \nu_\tau)_{\rm SM}$ if the Type~II 2HDM is adopted. That situation, however, is not of interest for our purpose since our scan does not allow 
$\tan \beta >15$. A plot of the resulting  $B(B\to \tau \nu_\tau)_{\rm 2HDM}$ as a function of $\tan\beta$ is shown in Fig.~\ref{fig:3} for the extreme values of charged Higgs boson, $m_{H^\pm}=400$~GeV and $m_{H^\pm}=1$~TeV, and compared to the experimental value at $2\sigma$-level. 
In the case of $B(D_s\to \tau \nu_\tau)_{\rm 2HDM}$ all type of 2HDM remain perfectly consistent with the Standard Model prediction which agrees with the experimentally established $B(D_s\to \tau \nu_\tau)=5.54(24)\times 10^{-2}$~\cite{PDG}.

As for the semileptonic decay, we consider the ratio $R_D = B(B\to D\tau \nu_\tau)/B(B\to D\mu \nu_\mu)$ in which a significant part of theoretical uncertainties cancel. Its value has been  measured in three experiments~\cite{Lees:2013uzd,Huschle:2015rga, RD_LHC} leading to an average of $R_D=0.41(5)$. That result is consistent with the SM value, $R_D^{\rm SM}=0.31(2)$~\cite{us}, at less than $2\sigma$-level and also with the 2HDM scenarios discussed here. Like in the case of leptonic decay, only in the Type II model at moderately large values of $\tan\beta$ one can see a small deviation with respect to the Standard Model, as shown in Fig.~\ref{fig:3}. That deviation is however too small to be probed experimentally for $\tan\beta < 15$.  
As for the loop induced processes, one extra constraint can be obtained from the comparison between the experimentally established ${B} (B_s \to \mu^+\mu^-)_{\mathrm{exp}} = \left(2.8^{+0.7}_{-0.6} \right)\times 10^{-9}$~\cite{CMS:2014xfa} with the Standard Model prediction~\cite{Bobeth:2013uxa} leading to 
\bea
{R}_{s\mu}\equiv\dfrac{B(B_s \to \mu^+\mu^-)}{B(B_s \to \mu^+\mu^-)_{\mathrm{SM}}}, \quad{\rm and}\quad {R}_{s\mu}^{\text{exp}}=0.76^{+0.20}_{-0.18}\,.
\eea
Using, instead of the Standard Model, the expressions for Wilson coefficients computed in a generic 2HDM~\cite{Li:2014fea} leads to the exclusion bound on the very low $\tan\beta \lesssim 0.7$, as shown in Fig.~\ref{fig:3}.

\section{Conclusion}

In this paper we showed that the CP-odd Higgs is a plausible candidate for the
resonance observed by both the CMS and the ATLAS experiments at LHC in the
diphoton spectrum around $750$~GeV.  From the general considerations in the
framework of 2HDM, and after fixing $m_h=125.7(4)$~GeV and $m_A=750(30)$~GeV,
we find the upper and lower bounds to the masses of the remaining two Higgs
bosons, namely \bea 400\ \gev \lesssim m_{H^\pm}\lesssim 1~{\rm TeV}, \qquad
200\ \gev \lesssim m_{H}\lesssim 1~{\rm TeV}.  \eea From our scan, in which we
used the general constraints spelled out in
Eqs.~(\ref{contrainte}--\ref{pertunit}), we also
find that the preferred values of $\tan\beta$ are relatively small, $\tan\beta
< 15$, with most of the points concentrated in the region $\tan\beta < 5$. The
width of the CP-odd Higgs is dominated by the $A\to t\bar t$ mode and its
value significantly depends on $\tan\beta$. We find that the width can be
large [as large as $\Gamma(A)\simeq 40$~GeV, for $\tan\beta =1$]. Furthermore,
we find that $|\cos(\beta-\alpha)|\lesssim 0.3$, i.e. not far from the
Standard Model, and in agreement with the results of the SFitter
analysis~\cite{Corbett:2015ksa}, and with the direct searches~\cite{lydia,others}.

We then checked that for the range of $\tan\beta$ and $m_{H^\pm}$ obtained
from our scan, the semileptonic and leptonic decay modes are not significantly
modified with respect to the Standard Model predictions. We also checked that the
spectrum of the 2HDM considered here is fully consistent with electroweak
precision data encoded in the $S$, $T$, $U$ parameters.

\begin{figure}[t!]
\centering
\includegraphics[width=0.5\linewidth]{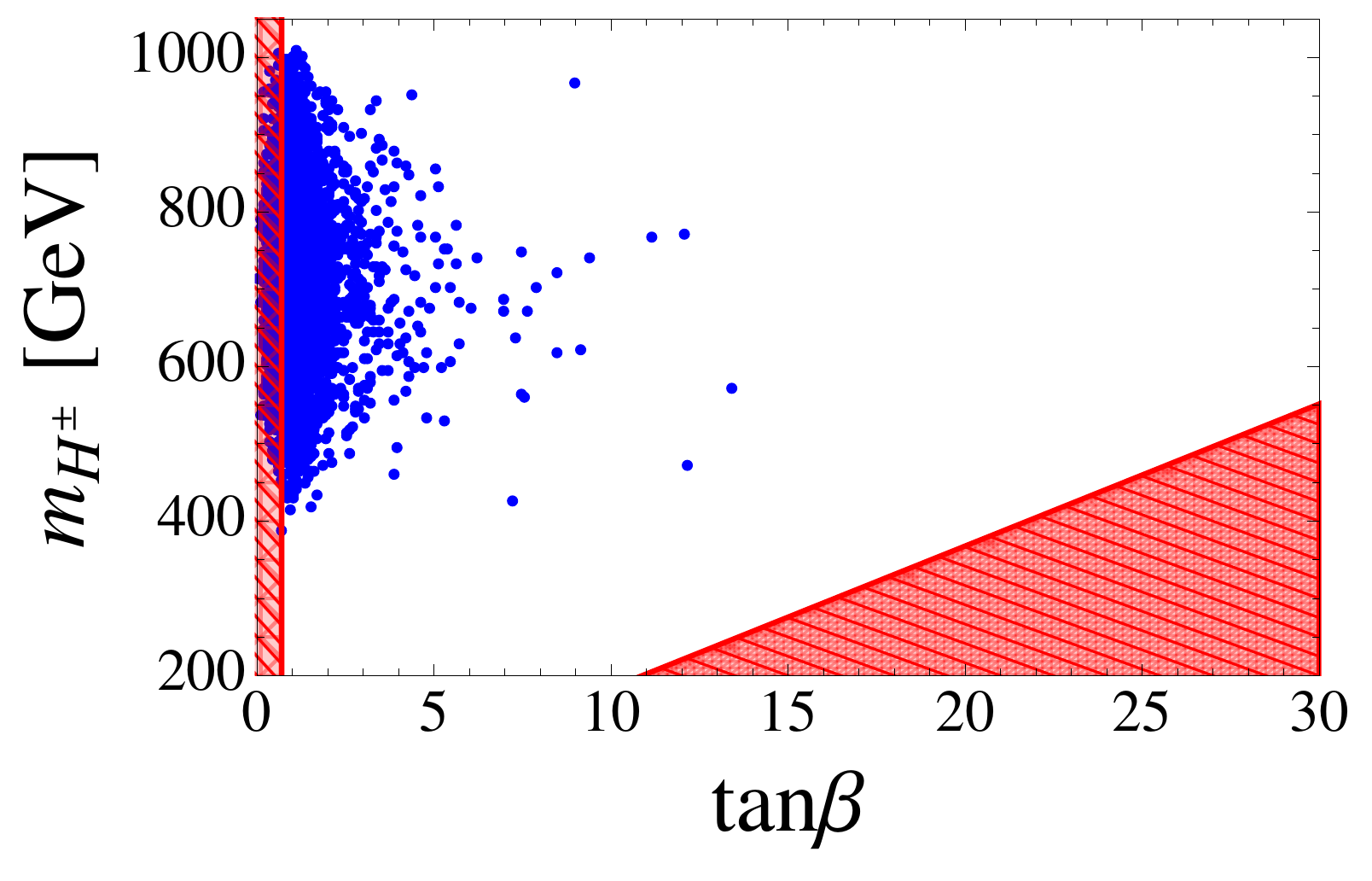}~\includegraphics[width=0.5\linewidth]{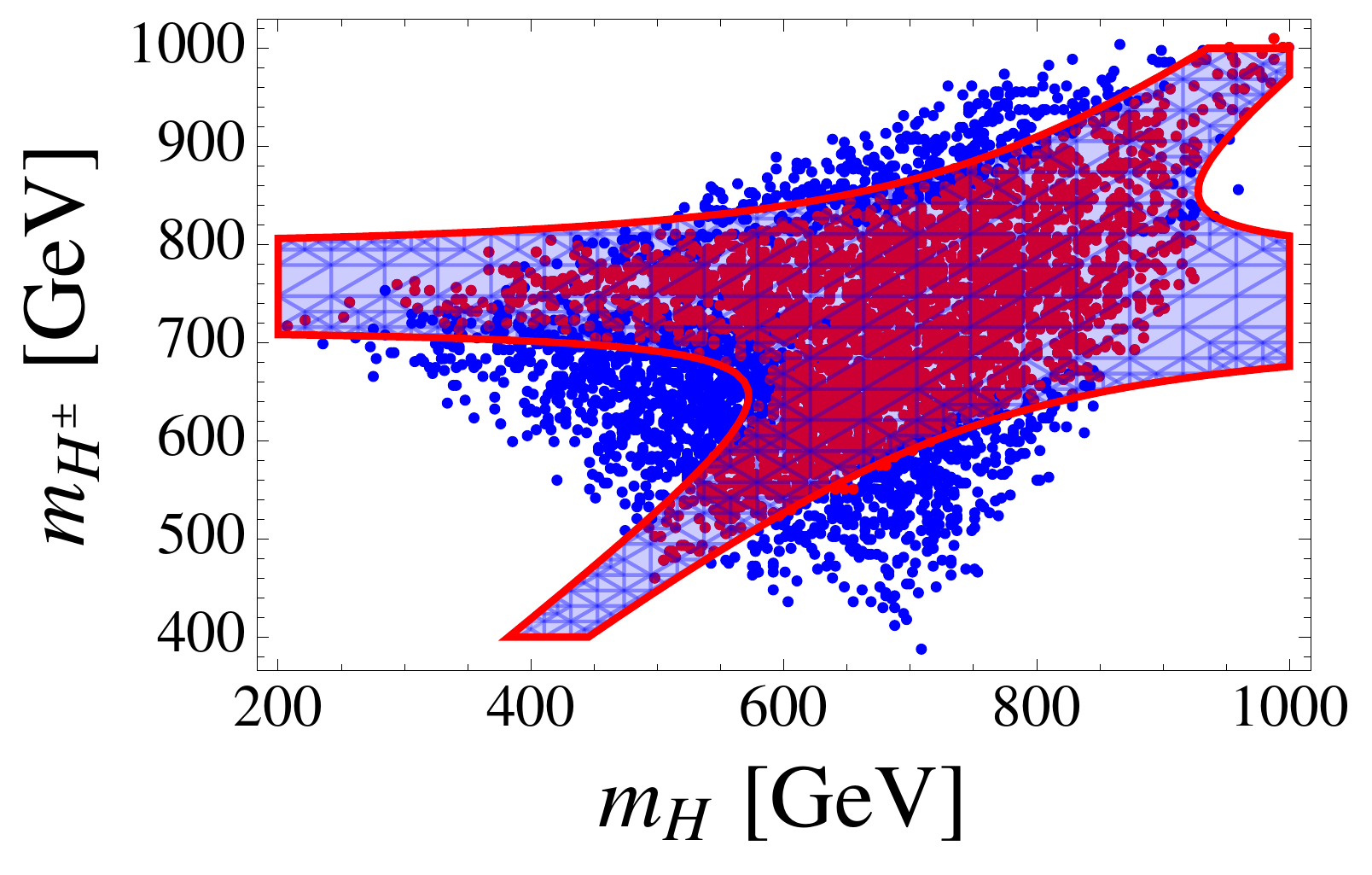}
\caption{\small\sl In the left plot we superpose the results of Sec.~\ref{sec:flavor} and the scan of allowed points presented in Fig.~\ref{fig:1}: very low $\tan\beta$ are forbidden by $B(B_s\to \mu\mu)$ whereas the constraint from $B\to\tau\nu$ is model dependent and in Type~II 2HDM it results in eliminating the large values of $\tan\beta$ for lower $m_{H^\pm}$, region already excluded by our scan made in Sec.~\ref{sec:2}.  In the right plot we superpose the results of Sec.~\ref{sec:EW} and the plot presented in Fig.~\ref{fig:1}:  we see that the electroweak precision data further restrict the region of allowed masses although, broadly speaking, the bounds we derived in Sec.~\ref{sec:2} remains unchanged.}
\label{fig:5}
\end{figure}

The plots, in which all the constraints considered in this paper are included, are presented in Fig.~\ref{fig:5}. Notice in particular that 
$B(B\to \tau \nu_\tau)$ and $R_D$ provide very similar bounds in the $(\tan\beta,m_{H^\pm})$ plane for 
$m_{H^\pm}\in (0.4,1)$~TeV. They, at best, exclude the large values of $\tan\beta$ in the Type~II model, otherwise they are insensitive to 
the parameter space (small $\tan\beta$) we are considering here. $B(B_s\to \mu^+\mu^-)$, instead, provides an important constraint, i.e. exclusion 
of the very small values of $\tan\beta$. Since that last constraint involves coupling to the top-quark, it is independent of the type of the 2HDM. 

Finally, we need to stress once again that our ambition was not to provide a full scenario of physics beyond the Standard Model. Rather, we merely 
attempt whether or not the recent experimental hint of the excess at LHC can be consistent with the interpretation of the CP-odd Higgs in the general framework of 2HDM.  
We find that scenario plausible and the repercussions on the remaining Higgs bosons look quite appealing because they can be either confirmed or 
refuted experimentally quite soon. Since the announcement of the LHC results~\cite{LHCnew} many authors discussed  
$\sigma(gg\to X)B(X\to \gamma\gamma)$, where $X$ stands for the resonance we
claim to be consistent with the CP-odd Higgs, being larger than expected in the
2HDM alone. At this point one should be careful about
interpretation of such results because: (a) they are not published or publicly announced, and
(b) a careful study of the signal strength and of the signal-background interference, including the appropriate cuts, is 
mandatory.
The conclusion we reached here is
solely based on considerations of the spectrum of scalars and it cannot be
significantly changed in the presence of additional fermionic degrees of
freedom, assuming the mixing between the Standard Model fermions and the extra heavy fermions is indeed small.

\section*{Acknowledgments} 
This work was partially supported by Funda\c{c}\~ao de Amparo \`a
Pesquisa do Estado de S\~ao Paulo (FAPESP) and Conselho Nacional de
Ci\^encia e Tecnologia (CNPq). D.B. acknow\-led\-ges the INP-CNRS support via{\sl ``Inphyniti"}.


\begin{thebibliography}{99}


\bibitem{PDG}
  K.~A.~Olive {\it et al.}  [Particle Data Group Collaboration],
  Chin.\ Phys.\ C {\bf 38} (2014) 090001.


\bibitem{LHCnew} LHC seminar {\sl ``ATLAS and CMS physics results from Run 2"}, talks by Jim Olsen and Marumi Kado,
CERN, 15 Dec. 2015.; ATLAS note, ATLAS-CONF-2015-081; CMS note, CMS PAS EXO-15-004.

 
 
\bibitem{Franceschini:2015kwy}
  R.~Franceschini {\it et al.},
  arXiv:1512.04933 [hep-ph].


\bibitem{Harigaya:2015ezk}
  K.~Harigaya and Y.~Nomura,
  arXiv:1512.04850 [hep-ph];
  Y.~Nakai, R.~Sato and K.~Tobioka,
  arXiv:1512.04924 [hep-ph].


\bibitem{Higaki:2015jag}
  T.~Higaki, K.~S.~Jeong, N.~Kitajima and F.~Takahashi,
  arXiv:1512.05295 [hep-ph];


\bibitem{Buttazzo:2015txu}
  D.~Buttazzo, A.~Greljo and D.~Marzocca,
  arXiv:1512.04929 [hep-ph];
  S.~D.~McDermott, P.~Meade and H.~Ramani,
  arXiv:1512.05326 [hep-ph].


\bibitem{Ellis:2015oso}
  J.~Ellis, S.~A.~R.~Ellis, J.~Quevillon, V.~Sanz and T.~You,
  arXiv:1512.05327 [hep-ph].


\bibitem{Angelescu:2015uiz}
  A.~Angelescu, A.~Djouadi and G.~Moreau,
  arXiv:1512.04921 [hep-ph];
  M.~Low, A.~Tesi and L.~T.~Wang,
  arXiv:1512.05328 [hep-ph].


\bibitem{Bellazzini:2015nxw}
  B.~Bellazzini, R.~Franceschini, F.~Sala and J.~Serra,
  arXiv:1512.05330 [hep-ph];
 C.~Petersson and R.~Torre,
  arXiv:1512.05333 [hep-ph].


\bibitem{DiChiara:2015vdm}
  S.~Di Chiara, L.~Marzola and M.~Raidal,
  arXiv:1512.04939 [hep-ph];
  R.~S.~Gupta, S.~Jäger, Y.~Kats, G.~Perez and E.~Stamou,
  arXiv:1512.05332 [hep-ph].



\bibitem{Knapen:2015dap}
  S.~Knapen, T.~Melia, M.~Papucci and K.~Zurek,
  arXiv:1512.04928 [hep-ph].


\bibitem{Mambrini:2015wyu}
  Y.~Mambrini, G.~Arcadi and A.~Djouadi,
  arXiv:1512.04913 [hep-ph];
 M.~Backovic, A.~Mariotti and D.~Redigolo,
  arXiv:1512.04917 [hep-ph].

\bibitem{Dev:2014yca}
  P.~S.~Bhupal Dev and A.~Pilaftsis,
  JHEP {\bf 1412} (2014) 024
   [JHEP {\bf 1511} (2015) 147]
  [arXiv:1408.3405 [hep-ph]].



\bibitem{Greiner:2015ixr}
  N.~Greiner, S.~Liebler and G.~Weiglein,
  arXiv:1512.07232 [hep-ph].
  
\bibitem{Branco:2011iw}
  G.~C.~Branco, P.~M.~Ferreira, L.~Lavoura, M.~N.~Rebelo, M.~Sher and J.~P.~Silva,
  Phys.\ Rept.\  {\bf 516} (2012) 1
  [arXiv:1106.0034 [hep-ph]].


\bibitem{Kanemura:2004mg}
  S.~Kanemura, Y.~Okada, E.~Senaha and C.-P.~Yuan,
  Phys.\ Rev.\ D {\bf 70} (2004) 115002
  [hep-ph/0408364].

\bibitem{Gunion:2002zf}
  J.~F.~Gunion and H.~E.~Haber,
  Phys.\ Rev.\ D {\bf 67} (2003) 075019
  [hep-ph/0207010].

\bibitem{Barroso:2013awa}
  A.~Barroso, P.~M.~Ferreira, I.~P.~Ivanov and R.~Santos,
  JHEP {\bf 1306} (2013) 045
  [arXiv:1303.5098 [hep-ph]].


\bibitem{unitarity}
 S.~Kanemura, T.~Kubota and E.~Takasugi,
  Phys.\ Lett.\ B {\bf 313} (1993) 155
  [hep-ph/9303263];
 B.~Swiezewska,
  Phys.\ Rev.\ D {\bf 88} (2013) 5,  055027
   [ibid {\bf 88} (2013) 11,  119903]
  [arXiv:1209.5725 [hep-ph]].





\bibitem{Corbett:2015ksa}
  T.~Corbett, O.~J.~P.~Eboli, D.~Goncalves, J.~Gonzalez-Fraile, T.~Plehn and M.~Rauch,
  JHEP {\bf 1508} (2015) 156
  [arXiv:1505.05516 [hep-ph]].

\bibitem{lydia}
L.~Brenner, {\sl ``Constraints on new phenomena through Higgs coupling measurements with the ATLAS detector"}, talk presented at the $23^{rd}$ International Conference on Supersymmetry and Unification of 
Fundamental Interactions, SUSY 2015, c.f https://indico.cern.ch/event/331032/session/7/contribution/237

\bibitem{others}
 G.~Aad {\it et al.} [ATLAS Collaboration],
  JHEP {\bf 1503} (2015) 088
  [arXiv:1412.6663 [hep-ex]];
  Phys.\ Rev.\ D {\bf 92} (2015) 5,  052002
  [arXiv:1505.01609 [hep-ex]];
  arXiv:1512.03704 [hep-ex];
V.~Khachatryan {\it et al.} [CMS Collaboration],
  JHEP {\bf 1511} (2015) 018
  [arXiv:1508.07774 [hep-ex]];
CMS Collaboration [CMS Collaboration],
  CMS-PAS-HIG-14-029;
P.~Onyisi [ATLAS and CMS Collaborations],
  PoS FPCP {\bf 2015} (2015) 023.




\bibitem{Crivellin:2013wna}
  A.~Crivellin, A.~Kokulu and C.~Greub,
  Phys.\ Rev.\ D {\bf 87} (2013) 9,  094031
  [arXiv:1303.5877 [hep-ph]];
ibid {\bf 86} (2012) 054014
  [arXiv:1206.2634 [hep-ph]];
 F.~Mahmoudi and O.~Stal,
  Phys.\ Rev.\ D {\bf 81} (2010) 035016
  [arXiv:0907.1791 [hep-ph]].



\bibitem{mfv}
  G.~D'Ambrosio, G.~F.~Giudice, G.~Isidori and A.~Strumia,
  Nucl.\ Phys.\ B {\bf 645} (2002) 155
  [hep-ph/0207036];
G.~Blankenburg and G.~Isidori,
  Eur.\ Phys.\ J.\ Plus {\bf 127} (2012) 85
  [arXiv:1107.1216 [hep-ph]];
E.~Nikolidakis and C.~Smith,
  Phys.\ Rev.\ D {\bf 77} (2008) 015021
  [arXiv:0710.3129 [hep-ph]].

\bibitem{CONSTRAINTS}
  A.~Celis, V.~Ilisie and A.~Pich,
  JHEP {\bf 1307} (2013) 053
  [arXiv:1302.4022 [hep-ph]];
W.~Altmannshofer, S.~Gori and G.~D.~Kribs,
  Phys.\ Rev.\ D {\bf 86} (2012) 115009
  [arXiv:1210.2465 [hep-ph]].


\bibitem{Buras:2010mh}
  A.~J.~Buras, M.~V.~Carlucci, S.~Gori and G.~Isidori,
  JHEP {\bf 1010} (2010) 009
  [arXiv:1005.5310 [hep-ph]].


\bibitem{pich}
M.~Jung, A.~Pich and P.~Tuzon,
  JHEP {\bf 1011} (2010) 003
  [arXiv:1006.0470 [hep-ph]];
A.~Pich and P.~Tuzon,
  Phys.\ Rev.\ D {\bf 80} (2009) 091702
  [arXiv:0908.1554 [hep-ph]].


\bibitem{Spira:1995rr}
  M.~Spira, A.~Djouadi, D.~Graudenz and P.~M.~Zerwas,
  Nucl.\ Phys.\ B {\bf 453} (1995) 17
  [hep-ph/9504378].


\bibitem{Gunion:1991cw}
  J.~F.~Gunion, H.~E.~Haber and C.~Kao,
  Phys.\ Rev.\ D {\bf 46} (1992) 2907.



\bibitem{Barbieri:2006bg}
  R.~Barbieri, L.~J.~Hall, Y.~Nomura and V.~S.~Rychkov,
  Phys.\ Rev.\ D {\bf 75} (2007) 035007
  [hep-ph/0607332].

\bibitem{Baak:2014ora}
  M.~Baak {\it et al.} [Gfitter Group Collaboration],
  Eur.\ Phys.\ J.\ C {\bf 74} (2014) 3046
  [arXiv:1407.3792 [hep-ph]].


\bibitem{FLAG}
 S.~Aoki {\it et al.},
  Eur.\ Phys.\ J.\ C {\bf 74} (2014) 2890
  [arXiv:1310.8555 [hep-lat]].



\bibitem{Lees:2012ju}
  J.~P.~Lees {\it et al.} [BaBar Collaboration],
  Phys.\ Rev.\ D {\bf 88} (2013) 3,  031102
  [arXiv:1207.0698 [hep-ex]].


\bibitem{Kronenbitter:2015kls}
  B.~Kronenbitter {\it et al.} [Belle Collaboration],
  Phys.\ Rev.\ D {\bf 92} (2015) 5,  051102
  [arXiv:1503.05613 [hep-ex]].



\bibitem{CKMology}
  J.~Charles {\it et al.},
  Phys.\ Rev.\ D {\bf 91} (2015) 7,  073007
  [arXiv:1501.05013 [hep-ph]];
M.~Bona {\it et al.} [UTfit Collaboration],
  Phys.\ Lett.\ B {\bf 687} (2010) 61
  [arXiv:0908.3470 [hep-ph]].



\bibitem{Lees:2013uzd}
  J.~P.~Lees {\it et al.} [BaBar Collaboration],
  Phys.\ Rev.\ D {\bf 88} (2013) 7,  072012
  [arXiv:1303.0571 [hep-ex]].

\bibitem{Huschle:2015rga}
  M.~Huschle {\it et al.} [Belle Collaboration],
  Phys.\ Rev.\ D {\bf 92} (2015) 7,  072014
  [arXiv:1507.03233 [hep-ex]].

\bibitem{RD_LHC} G. Ciezarek [LHCb Collaboration], talk presented at Flavor Physics \& CP violation 2015 (Nagoya, Japan,
25-29 May 2015), [fpcp2015.hepl.phys.nagoya-u.ac.jp].

\bibitem{us}
  D.~Becirevic, N.~Kosnik and A.~Tayduganov,
  Phys.\ Lett.\ B {\bf 716} (2012) 208
   [arXiv:1206.4977 [hep-ph]];
J.~A.~Bailey {\it et al.} [MILC Collaboration],
  Phys.\ Rev.\ D {\bf 92} (2015) 3,  034506
  [arXiv:1503.07237 [hep-lat]];
H.~Na {\it et al.} [HPQCD Collaboration],
  Phys.\ Rev.\ D {\bf 92} (2015) 5,  054510
  [arXiv:1505.03925 [hep-lat]];
M.~Atoui, V.~Morenas, D.~Becirevic and F.~Sanfilippo,
  Eur.\ Phys.\ J.\ C {\bf 74} (2014) 5,  2861
  [arXiv:1310.5238 [hep-lat]].



\bibitem{CMS:2014xfa}
  V.~Khachatryan {\it et al.} [CMS and LHCb Collaborations],
  Nature {\bf 522} (2015) 68
  [arXiv:1411.4413 [hep-ex]].

\bibitem{Bobeth:2013uxa}
  C.~Bobeth, M.~Gorbahn, T.~Hermann, M.~Misiak, E.~Stamou and M.~Steinhauser,
  Phys.\ Rev.\ Lett.\  {\bf 112} (2014) 101801
  [arXiv:1311.0903 [hep-ph]].


\bibitem{Li:2014fea}
  X.~Q.~Li, J.~Lu and A.~Pich,
  JHEP {\bf 1406} (2014) 022
  [arXiv:1404.5865 [hep-ph]].


\end{thebibliography}
\end{document}